\documentclass[prd,reprint,nofootinbib,superscriptaddress,preprintnumbers,showpacs]{revtex4-1}
\usepackage{amsfonts}
\usepackage{amssymb}
\usepackage{amsmath}
\usepackage{graphicx,color}
\usepackage{dcolumn}
\usepackage{epsfig}
\usepackage{bm}
\usepackage{bbm}
\usepackage{ulem}
\usepackage{hyperref}
\usepackage{lineno}
\usepackage{multirow}

\usepackage{color}

\def\gtap{\ \raise.3ex\hbox{$>$\kern-.75em\lower1ex\hbox{$\sim$}}\ }
\def\ltap{\ \raise.3ex\hbox{$<$\kern-.75em\lower1ex\hbox{$\sim$}}\ }
\begin{document}

\title{
Three-Body Unitary Coupled-Channel Analysis on $\eta(1405/1475)$
}
\author{S.X. Nakamura}
\email{satoshi@ustc.edu.cn}
\affiliation{
University of Science and Technology of China, Hefei 230026, 
China
}
\affiliation{
State Key Laboratory of Particle Detection and Electronics (IHEP-USTC), Hefei 230036, China}
\author{Q. Huang}
\affiliation{
Department of Physics, Nanjing Normal University, Nanjing, Jiangsu 210097, China
}
\author{J.-J. Wu}
\email{wujiajun@ucas.ac.cn}
\affiliation{
School of Physical Sciences, University of Chinese Academy of Sciences (UCAS), Beijing 100049, China
}
\author{H.P. Peng}
\author{Y. Zhang}
\author{Y.C. Zhu}
\affiliation{
University of Science and Technology of China, Hefei 230026, 
China
}
\affiliation{
State Key Laboratory of Particle Detection and Electronics (IHEP-USTC), Hefei 230036, China}

\begin{abstract}
The recent BESIII data on $J/\psi\to\gamma(K_SK_S\pi^0)$,
which is significantly more precise than earlier 
$\eta(1405/1475)$-related
data, enables quantitative discussions on $\eta(1405/1475)$ at the previously unreachable level.
We conduct a three-body unitary coupled-channel analysis of experimental Monte-Carlo outputs for radiative $J/\psi$ decays via $\eta(1405/1475)$: 
$K_SK_S\pi^0$ Dalitz plot distributions from the BESIII, and branching ratios of $\gamma(\eta\pi^+\pi^-)$ and $\gamma(\gamma\pi^+\pi^-)$ final states relative to that of $\gamma(K\bar{K}\pi)$.
Our model systematically considers (multi-)loop diagrams and an associated triangle singularity, which is critical in making excellent predictions on $\eta(1405/1475)\to \pi\pi\pi$ lineshapes and branching ratios.
The $\eta(1405/1475)$ pole locations are revealed for the first time.
Two poles for $\eta(1405)$ are found on different Riemann sheets of the
$K^*\bar{K}$ channel, while one pole for $\eta(1475)$.
The $\eta(1405/1475)$ states are described with two bare states dressed by 
continuum states.
The lower bare state would be an excited $\eta^\prime$, while the higher one could be an excited $\eta^{(\prime)}$, hybrid, glueball, or their mixture.
This work presents the first-ever pole determination based on a
manifestly three-body unitary coupled-channel framework applied to
experimental three-body final state distributions (Dalitz plots). 
\end{abstract}

\maketitle


\section{Introduction}
The nature of isoscalar pseudoscalar meson(s) in 1.4--1.5~GeV region, $\eta(1405/1475)$, has been controversial.
On the experimental side, two different states seem to work for $K\bar{K}\pi$ final states produced in $\pi^-p$ scattering~\cite{e852,e769}, $p\bar{p}$ annihilations~\cite{obelix2002},
and radiative $J/\psi$ decays~\cite{mark3_1990,dm2_1992}.
However, only one resonant peak, whose position is somewhat process-dependent, is observed in:
$\eta\pi\pi$ final states in $p\bar{p}$ annihilation~\cite{amsler} and $J/\psi$ decays accompanied by $\gamma$~\cite{mark3_jpsi-gamma-eta-pipi,bes_jpsi-gamma-eta-pipi,dm2_jpsi-gamma-eta-pipi} and $\omega$~\cite{bes3_jpsi-omega-eta-pipi};
$K\bar{K}\pi$ and $\eta\pi\pi$ final states in $\gamma\gamma$ collisions~\cite{L3};
$\gamma\rho^0$ final states in radiative $J/\psi$ decays~\cite{bes2_rhog,mark3_rhog,dm2_jpsi-gamma-eta-pipi} and $p\bar{p}$ annihilation~\cite{amsler}.
These data are statistically limited, allowing various theoretical descriptions.
In particular, whether $\eta(1405/1475)$ is one or two states remains as a major puzzle.

The quark model predicts only one state, a radially excited $\eta'(958)$, in this energy region
and the ideal mixing ($s\bar{s}$)~\cite{pdg,barnes1997} seems consistent with a lattice QCD (LQCD)~\cite{dudek2013}.
To accommodate two states, $\eta(1405)$ was proposed to be a glueball~\cite{faddeev2004}, which however is disfavored by LQCD predicting a significantly heavier mass~\cite{bali1993,morningstar1999,chen2006,richards2010,chen2111}.
The $\eta(1405/1475)$ couples to quasi two-body channels such as $K^*\bar{K}$ and $a_0\pi$ that further decay to three-body channels such as $K\bar{K}\pi$ and $\pi\pi\eta$, forming a complicated coupled-channel system.
Also, a kinematical triangle singularity is caused by the coupled-channel dynamics and plays an important role~\cite{wu2012,wu2013,Aceti:2012dj,du2019}.  
Thus, a sophisticated coupled-channel analysis of high quality data has been long-awaited to pin  down the nature of $\eta(1405/1475)$.

The recent high-statistics BESIII experiment provides a precious opportunity to improve our understanding of $\eta(1405/1475)$.
They collected $\sim 10^{10} J/\psi$ decay samples, and conducted an amplitude analysis on 
$J/\psi\to\gamma(K_SK_S\pi^0)$~\cite{bes3_mc}.
Their bin-by-bin analysis of the $K_SK_S\pi^0$ invariant mass extracted a $J^{PC}=0^{-+}$ contribution. 
Then, their energy-dependent analysis identified two $\eta(1405/1475)$ states with a high statistical significance, and determined their Breit-Wigner (BW) masses and widths.
However, the BW amplitude does not respect the unitary and is therefore not suitable in situations where more than one resonances are overlapping and/or a resonance is close to its decay channel threshold~\cite{3pi-2};
the situations apply to $\eta(1405)$ and $\eta(1475)$ that are overlapping, and $\eta(1405)$ being close to the $K^*\bar{K}$ threshold.
Thus, an important issue is to determine the $\eta(1405/1475)$ pole locations, which can be achieved by analytically continuing a unitary coupled-channel $J/\psi$ decay amplitude fitted to the BESIII data.

Another puzzling issue is a large isospin-violation in $\eta(1405/1475)\to\pi\pi\pi$~\cite{bes3-3pi}.
An explanation has been proposed in Refs.~\cite{wu2012,wu2013,Aceti:2012dj}:
the $K^*\bar{K}K$-loop mechanism involving a triangle singularity causes the large isospin-violation due to the mass difference between $K^\pm$ and $K^0$.
Now the issue is to confirm this explanation by examining whether the three-body unitary coupled-channel model fitted to the recent BESIII data~\cite{bes3_mc} can also consistently describe lineshapes and branchings of the three-pion final states; 
the triangle singularity mechanism is automatically included in the unitary framework.

\begin{figure*}
\begin{center}
\includegraphics[width=0.95\textwidth]{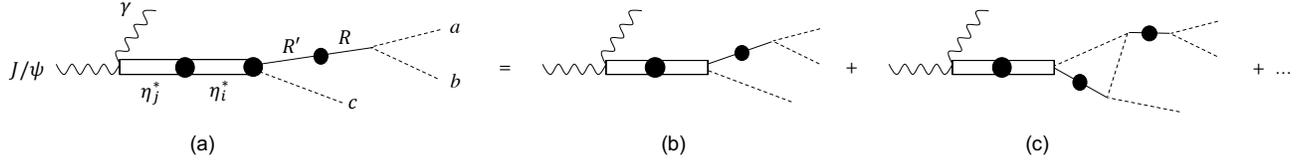}
\end{center}
 \caption{(a) Radiative $J/\psi$ decay model.
The dashed and solid
lines represent pseudoscalar mesons and
bare two-meson resonances $R$, respectively.
 The double lines represent bare $\eta(1405/1475)$ states. 
Dressed propagators and vertices are represented by 
the solid circles.
Main $\eta^*$ decay mechanisms are
(b) direct decays and (c) single triangle mechanisms.
 }
\label{fig:diag}
\end{figure*}

In this work~\footnote{A fuller account of this work will be given separately~\cite{follow-up}},
we conduct a coupled-channel analysis of radiative $J/\psi$ decays via $\eta(1405/1475)$.
Our three-body unitary coupled-channel model is fitted to $K_SK_S\pi^0$ Dalitz plot distributions from the BESIII Monte-Carlo (MC) simulation~\cite{bes3_mc}, as well as to branching fractions of $\eta\pi^+\pi^-$ and $\rho^0\gamma$ relative to that of $K\bar{K}\pi$.
The model will clarify main $\eta(1405/1475)$ decay mechanisms and predict 
$\eta(1405/1475)\to \eta\pi\pi, \pi\pi\pi$ lineshapes and branchings. 
The analysis leads to answering three major puzzles regarding $\eta(1405/1475)$:
\begin{itemize}
  \setlength{\itemsep}{0mm}
 \item Process-dependent lineshapes of $\eta(1405/1475)$ decays 
 \item Large isospin-violation in $\eta(1405/1475)\to \pi\pi\pi$
 \item One or two states of $\eta(1405/1475)$, equivalently, pole structure
\end{itemize}


\section{Model}

Our three-body unitary coupled-channel model is primarily based on the formulation
in Refs.~\cite{3pi,d-decay}.
Similar three-body unitary formulations were presented recently~\cite{gwu1,jpac1,jpac2}.
A notable extension here is to consider charge-dependent particle masses for
describing the isospin-violations.
Thus, a radiative $J/\psi$ decay amplitude\footnote{We denote a particle $x$'s mass, momentum, energy, spin, and $z$-component in the $abc$ center-of-mass frame by $m_x$, $\bm{p}_x$, $E_x$, $s_x$, and $s_x^z$,
respectively; $E_x=\sqrt{m_x^2+|\bm{p}_x|^2}$. The mass values are from Ref.~\cite{pdg}.}
via $\eta(1405/1475)$ excitations 
 is diagrammatically represented in
 Fig.~\ref{fig:diag}(a) and given by
\begin{eqnarray}
 A_{\gamma abc,J/\psi} &=& 
\sum^{\rm cyclic}_{abc}
\sum_{RR's_R^z}
\sum_{ij}
\Gamma_{ab,R}\,
\tau_{R,R'}(p_c,E-E_{c})\,
\nonumber\\
&&\times \bar{\Gamma}_{cR',\eta^*_i}(\bm{p}_c, E)\,
\bar{G}_{ij}(E)\,
\Gamma_{\gamma \eta^*_j,J/\psi}\, ,
\label{eq:amp_a}
\end{eqnarray}
where $a$, $b$, and $c$ are pseudoscalar mesons ($\pi$, $K$, $\eta$),
and $R$ denotes a two-meson resonance such as $K^*$, 
$K^*_0(700) (=\kappa)$,
$a_0(980)$, and $f_0(980)$;
cyclic permutations $(abc), (cab), (bca)$ are indicated by $\sum^{\text{cyclic}}_{abc}$;
the indices $i$ and $j$ specify one of bare $\eta^*$ states;
$E$ denotes the $abc$ total energy in the $abc$ center-of-mass (CM) frame.
We introduced 
a $J/\psi\to \gamma \eta^*_i$ vertex ($\Gamma_{\gamma\eta^*_j,J/\psi}$),
a dressed $\eta^*$ propagator ($\bar{G}_{ij}$),
a dressed $\eta^*_i\to Rc$ vertex ($\bar{\Gamma}_{cR,\eta^*_i}$),
a dressed $R$ propagator ($\tau_{R,R'}$), and
a $R\to ab$ vertex ($\Gamma_{ab,R}$).

The dressed $R$ propagator matrix is
\begin{eqnarray}
[\tau^{-1}(p,E) ]_{R,R'} &=& [E - E_{R}(p) ]\delta_{R,R'} - [\Sigma(p, E)]_{R,R'},
\label{eq:green-Rc}
\end{eqnarray}
where a matrix $\Sigma_{R,R'}$ is the $R$ self-energy caused by $\Gamma_{ab,R}$.
The dressed $\eta^*_i\to Rc$ vertices are
\begin{eqnarray}
\label{eq:dressed-g}
\bar \Gamma_{cR ,\eta^*_i}(\bm{p}_c ,E)&=& \!\int\! d^3q\,\Phi_{cR,c'R'}(\bm{p}_c,\bm{q};E)
\Gamma_{c'R',\eta^*_i}(\bm{q}) ,
\end{eqnarray}
with $\sum_{c'R's_{R'}^z}$ being implicit;
$\Phi=(1-\int d^3q\,V\tau)^{-1}$ is a wave function;
${\Gamma}_{cR,\eta^*_i}$ is a bare $\eta^*_i\to Rc$ vertex. 
The $Rc\to R'c'$ interaction $V$ includes $Z$-diagrams in which 
$R\to c'\bar{c}$ is followed by $\bar{c}c\to R'$ via a $\bar{c}$-exchange.
An isospin-violating $K^*\bar{K}\to f_0\pi$ process is caused by a $K$-exchange $Z$-diagram and $m_{K^\pm}\ne m_{K^0}$.
Formulas for the $Z$-diagrams can be found in Appendix~C of Ref.~\cite{3pi}.
Also, $V$ includes vector-meson exchange mechanisms, based on the hidden local symmetry model~\cite{hls}, for $K^*\bar{K}\leftrightarrow K^*\bar{K}, \bar{K}^*K$; see Appendix~A of Ref.~\cite{d-decay} for formulas.
The nonperturbative treatment of $V\tau$ in Eq.~(\ref{eq:dressed-g}) is a requirement from the three-body unitarity.

The dressed $\eta^*$ propagator is
\begin{eqnarray}
\left[\bar{G}^{-1}(E)\right]_{ij} = (E- m_{\eta^*_i})\delta_{ij} - \left[\Sigma_{\eta^*}(E)\right]_{ij}\,,
\label{eq:mstar-g1}
\end{eqnarray}
where $m_{\eta^*_i}$ is the bare mass and the $\eta^*$ self energy is
\begin{eqnarray}
[\Sigma_{\eta^*}(E)]_{ij} &=& 
\sum_{cRR's_R^z} 
\int d^3q\,
\Gamma_{cR ,\eta^*_i}(\bm{q})
\nonumber\\
&&\times 
\tau_{R,R'}(q,E-E_c(q)) 
\bar \Gamma_{cR' ,\eta^*_j}(\bm{q} ,E)  .
\label{eq:mstar-sigma}
\end{eqnarray}

The coupled-channels included in our default model are two bare $\eta^*$
states and $Rc=K^*(892)\bar K$, $\kappa\bar K$, $a_0(980)\pi$,
$a_2(1320)\pi$, $f_0\eta$, $\rho(770)\rho(770)$~\footnote{The $\rho\rho$
channel needs slight modifications of the presented formulas; see Ref.~\cite{follow-up}.},
and $f_0\pi$:
$\bar{K}^*(892) K$ and $\bar{\kappa} K$
are implicitly included to form positive $C$-parity states.
The bare $R$ states and their decay channels (two-meson continuum states) couple nonperturbatively to generate scattering amplitudes and resonance poles. 
Thus we can fix the coupling and cutoff parameters in $\Gamma_{ab,R}$ and $m_R$ (bare mass) by fitting $ab \to ab$ scattering data.
Meanwhile, 
real and complex coupling parameters in
${\Gamma}_{cR,\eta^*_i}$ and $\Gamma_{\gamma \eta^*_i,J/\psi}$,
respectively, 
and $m_{\eta^*_i}$ 
are fitted
to MC outputs for $J/\psi\to\gamma\eta(1405/1475)\to\gamma(abc)$ 
as detailed in Sec.~\ref{subsec:fit};
cutoffs of dipole form factors in 
${\Gamma}_{cR,\eta^*_i}$ are fixed to 700~MeV.
To describe the $\gamma(\pi^+\pi^-\gamma)$ final state, we assume the vector-meson dominance mechanism
where $\rho\rho$ from the dressed $\eta^*_i\to\rho\rho$ is followed by $\rho\to\gamma$ and $\rho\to\pi^+\pi^-$; no additional parameters.
We totally have 25 fitting parameters. 


\section{Results}
\subsection{Fit and comparison with data}
\label{subsec:fit}

Using the $J^{PC}=0^{-+}$ partial wave amplitude 
from the BESIII MC ($E$-dependent solution)
for $J/\psi\to\gamma K_SK_S\pi^0$~\cite{bes3_mc},
we generate $K_SK_S\pi^0$ Dalitz plot pseudodata 
for each of 30 $E$ bins (10~MeV bin width; labeled by $l$) in the range of $1300\le E\le 1600$~MeV.
The pseudodata is thus
detection efficiency-corrected and background-free.
The Dalitz plot for an $l$-th $E$ bin
is further binned 
by equally dividing 
$(0.95\, {\rm GeV})^2\le m^2_{K_SK_S}\le (1.50\, {\rm GeV})^2$ and 
$(0.60\, {\rm GeV})^2\le m^2_{K_S\pi^0}\le (1.15\, {\rm GeV})^2$ 
into $50\times 50$ bins (labeled by $m$); $m_{ab}$ is the $ab$ invariant mass.
The pseudodata includes $\sim 1.23 \times 10^5$ events in total, being consistent
with the BESIII data. 
The event numbers in
$\{l,m\}$ and $l$-th 
bins are 
$N_{l,m}$ and $\bar N_{l}(\equiv\sum_m N_{l,m})$, respectively, with
their statistical uncertainties 
$\sqrt{N_{l,m}}$ and $\sqrt{\bar N_{l}}$, respectively.
Fitting $\{N_{l,m}\}$ and $\{\bar N_{l}\}$ pseudodata would
constrain
the detailed decay dynamics and the 
resonant behavior (pole structure) of $\eta(1405/1475)$, respectively.
We generate and fit 50 pseudodata samples to estimate the statistical
uncertainty of the model with the bootstrap method~\cite{bootstrap}.

Ratios of partial decay widths are also fitted:
$R_1^{\rm exp}=
\Gamma  [J/\psi\to\gamma\eta(1405/1475)\to\gamma(K\bar{K}\pi)]/
\Gamma  [J/\psi\to\gamma\eta(1405/1475)\to\gamma(\eta\pi^+\pi^-)]
\sim 6.8-11.9$~\cite{pdg},
and 
$R_2^{\rm exp}
=\Gamma  [J/\psi\to\gamma\eta(1405/1475)\to\gamma(\rho^0\gamma)]/
\Gamma  [J/\psi\to\gamma\eta(1405/1475)\to\gamma(K\bar{K}\pi)]
= 0.015-0.043$~\cite{mark3_rhog,bes2_rhog}.
We calculate the partial widths $\Gamma$ by integrating the $E$ distributions for the $K\bar{K}\pi$, $\pi^+\pi^-\eta$, and $\pi^+\pi^-\gamma$ final states over the range of $1350$~MeV $<E< 1550$~MeV.
The above ratios can constrain parameters associated with the $f_0\eta$ and $\rho\rho$ channels that are not well determined by the $K_SK_S\pi^0$ Dalitz plots. 

\begin{figure}
\begin{center}
\includegraphics[width=.5\textwidth]{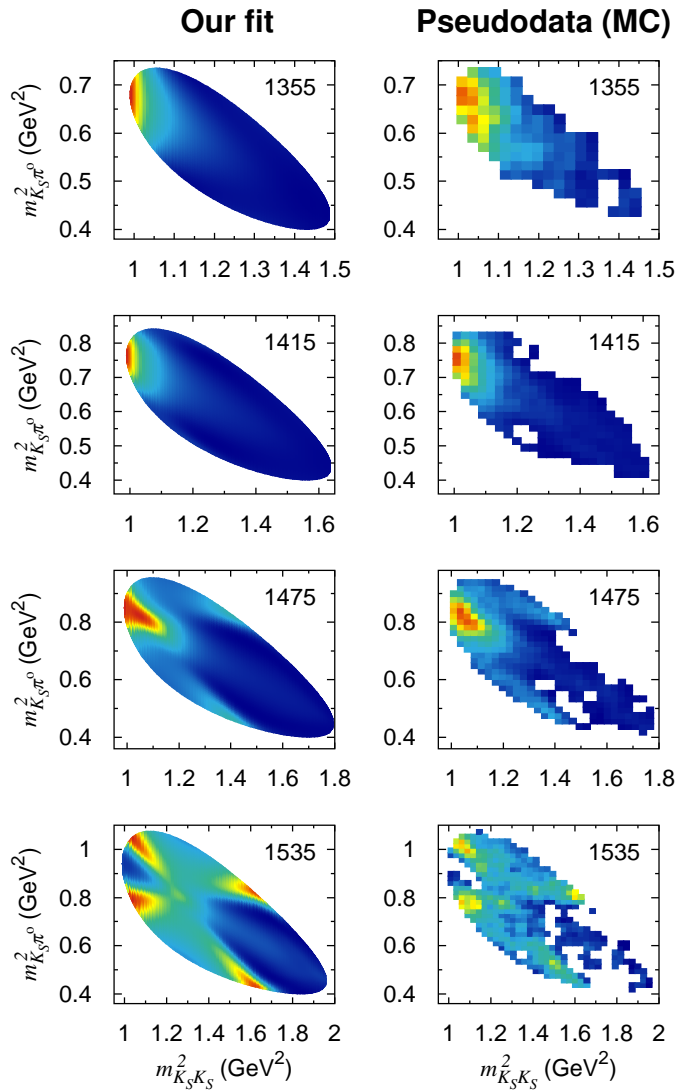}
\end{center}
 \caption{
Dalitz plot distributions of
$J/\psi\to\gamma\eta(1405/1475)\to\gamma(K_SK_S\pi^0)$.
The $E$ values (MeV) used in our calculation (central values of the MC $E$ bins)
are indicated.
 }
\label{fig:dalitz}
\end{figure}

Our default model is simultaneously fitted to the MC-based
$\{N_{l,m}\}$, $\{\bar N_{l}\}$, $R_1^{\rm exp}$, and $R_2^{\rm exp}$
with a $\chi^2$-minimization;
no direct fit to the actual BESIII data.
To keep a reasonable computational cost 
for calculating $\chi^2$ from $\{N_{l,m}\}$,
we compare $N_{l,m}$ with 
the differential decay width evaluated at the bin center and 
multiplied by the bin volume. 
Accordingly, $N_{l,m}$ on the phase-space boundary are omitted from 
the $\chi^2$ calculation. 
Also, a bin of $N_{l,m}<10$ is combined with neighboring bins
so that bins with more than 9 events go into
the $\chi^2$ calculation. 
The number of bins for $\{N_{l,m}\}$ is
4496--4575, depending on the pseudodata samples. 
$\chi^2$ from $\{\bar N_{l}\}$, $R_1^{\rm exp}$, and $R_2^{\rm exp}$
are appropriately weighted
so that these data can reasonably constrain the model. 
By fitting the 50 samples, we obtain
$\chi^2/{\rm ndf}=$ 1.40--1.54 (ndf: number of degrees of freedom)
from comparing with $\{N_{l,m}\}$, and
$R_1^{\rm th}\sim 7.5$ and $R_2^{\rm th}\sim 0.025$.

\begin{figure}
\begin{center}
\includegraphics[width=.48\textwidth]{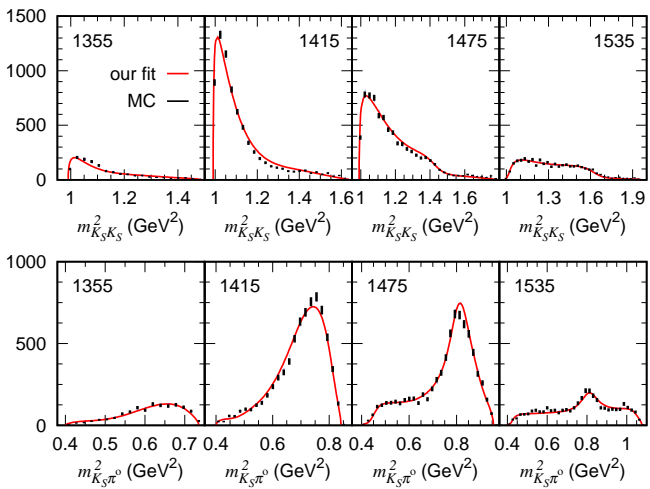}
\end{center}
 \caption{
The $K_SK_S$ (upper panels) and $K_S\pi^0$ (lower) invariant mass
 distributions (events/bin).
The $E$ values (MeV) are indicated.
 }
\label{fig:KK-Kpi}
\end{figure}

\begin{figure}[b]
\begin{center}
\includegraphics[width=.485\textwidth]{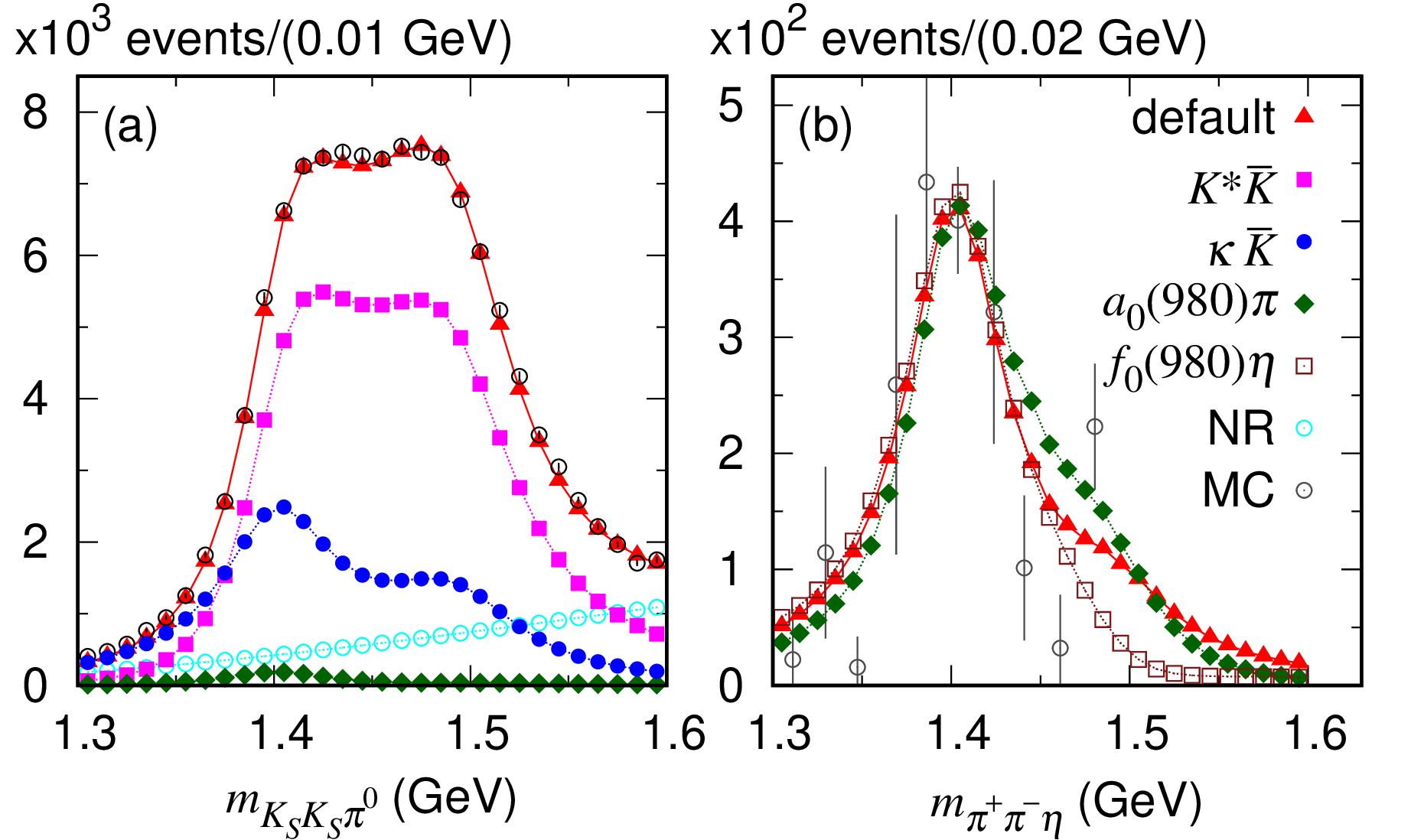}
\end{center}
 \caption{
The $E$ distributions for $J/\psi\to \gamma\eta(1405/1475)\to$ (a) $\gamma(K_SK_S\pi^0)$ and (b) $\gamma(\pi^+\pi^-\eta)$.
The default, final $Rc$, and nonresonant (NR) contributions, and 
MC outputs 
(\cite{bes3_mc} in (a), \cite{mark3_jpsi-gamma-eta-pipi} in (b))
are shown.
Lines connecting the points are just for guiding eyes.
 }
\label{fig:kkpi}
\end{figure}

In Fig.~\ref{fig:dalitz}, we show
the $K_SK_S\pi^0$ Dalitz plot distributions,
at representative $E$ values,
from one of the pseudodata samples 
and our default fit to them~\footnote{
Figures~\ref{fig:dalitz}(right), \ref{fig:KK-Kpi}, and \ref{fig:kkpi}(a) show the same 
pseudodata.}.
Their patterns agree well overall.
The $a_0(980)$-like peak is clearly seen 
near the $K_SK_S$ threshold for $1.3\ltap E\ltap 1.44$~GeV, while the $K^*$
peak is clear for $1.5\ltap E\ltap 1.6$~GeV.
The fits of a quite good quality are more clearly shown in Fig.~\ref{fig:KK-Kpi}
where the $a_0(980)$-like and $K^*(892)$ peak structures in the $K_SK_S$ and
$K_S\pi^0$ invariant mass distributions, respectively, are well
reproduced.
The absolute values of the distributions are large in the $\eta(1405/1475)$ peak region ($E=1.4-1.5$~GeV).
By integrating the distributions at each $E$, we obtain the $E$ distribution shown in Fig.~\ref{fig:kkpi}(a).

We show contributions from main $\eta^*\to K\bar{K}\pi$ decay mechanisms.
The $\eta^*$ decay mechanisms can be classified according to {\it final} $Rc$ states in Fig.~\ref{fig:diag}(a) that directly couple to the final $abc$ states.
As shown in Fig.~\ref{fig:kkpi}(a), the final $K^*\bar{K}$ and $\kappa\bar{K}$ give the first and second largest contributions, respectively. 
The clear $a_0(980)$-like peak in the $K_SK_S$ invariant mass spectra (Fig.~\ref{fig:KK-Kpi})
is mostly formed by a constructive interference within the
final Bose-symmetrized $K^*_SK_{S}\to \pi^0K_SK_S$ contribution at the $K_SK_S$ threshold; 
the small final $a_0(980)\pi$ contribution slightly sharpens the peak 
through an interference.

Our decay mechanisms are rather different from the BESIII MC~\cite{bes3_mc}
where the $a_0(980)\pi$ contribution is the largest overall, and 
the $K^*\bar{K}$ contribution is comparable only at $E\sim 1500$~MeV.
There are three important improvements in our model: including the
$\kappa\bar{K}$ channel,  fitting the ratio $R_1^{\rm exp}$, and
accounting for the coupled-channel effects.
The large $R_1^{\rm exp}$ is, albeit a large uncertainty, an important constraint on the final $a_0(980)\pi$ contribution to $K\bar{K}\pi$ since the coupling magnitude of $a_0(980)\to K\bar{K}$ relative to $a_0(980)\to \pi\eta$ is determined experimentally~\cite{a0_980_ppbar}.
Our $a_0(980)\pi$ contribution is small to fit $R_1^{\rm exp}$, and the final $\kappa\bar{K}$ contribution is significant.

Among the coupled-channel mechanisms included in Fig.~\ref{fig:diag}(a), direct decays [Fig.~\ref{fig:diag}(b)] and single triangle mechanisms [Fig.~\ref{fig:diag}(c)]
play an important role. 
The direct-decay and single-triangle mechanisms are dominant in the final
$K^*\bar{K}$ and $\kappa\bar{K}$ contributions, respectively, while 
they are comparable in the final $a_0(980)\pi$ contribution.
Figure~\ref{fig:kkpi}(a) shows that 
the broad peak structure from the full calculation is
mainly formed by
the final $K^*\bar{K}$ contribution.

To address whether $\eta(1405/1475)$ is one or two states,
we attempted to fit 
the BESIII MC output for $K_SK_S\pi^0$
with a single bare $\eta^*$ model.
The final $\kappa\bar{K}$ and $a_0\pi$ contributions have similar lineshapes peaking at $E\sim$1420~MeV, while the final $K^*\bar{K}$ contribution has a peak at 30--40~MeV higher
since $K^*\bar{K}$ is relatively $p$-wave and its threshold is at $E\sim$1400~MeV.
Their coherent sum cannot reproduce the $\sim$100~MeV wide flat peak, even if fitting the $E$ distribution only.
We thus conclude that two bare $\eta^*$ states are necessary to explain 
the $0^{-+}$ contribution of the BESIII MC.

Now our default model makes predictions for $\eta^*\to\pi\pi\pi$ and $\pi\pi\eta$.
The predicted $E$ dependence for $J/\psi\to\gamma\eta(1405/1475)\to\gamma(\eta\pi^+\pi^-)$
is shown in Fig.~\ref{fig:kkpi}(b).
The lineshape is consistent with the MC~\cite{mark3_jpsi-gamma-eta-pipi,bes_jpsi-gamma-eta-pipi}.
Although both $K\bar{K}\pi$ and $\pi\pi\eta$ originate from the same resonance(s),
the $\pi\pi\eta$ final states give a single peak at $m_{\pi\pi\eta}\sim$ 1400~MeV while
the $m_{K\bar{K}\pi}$ distribution has the broad peak. 
This is because $K\bar{K}\pi$ and $\pi\pi\eta$ are from different final $Rc$ states that
have different $E$ dependences. 
As shown in Fig.~\ref{fig:kkpi}(b) the comparable final $a_0(980)\pi$ 
and $f_0\eta$ contributions
explain the full result for the $\pi\pi\eta$ final states.
On the other hand, the $K\bar{K}\pi$ final states are mainly from the
final $K^*\bar{K}$ and $\kappa\bar{K}$ contributions, as seen in Fig.~\ref{fig:kkpi}(a).
This explains the process dependence of the
$\eta(1405/1475)$ lineshapes.

\begin{figure}
\begin{center}
\includegraphics[width=.49\textwidth]{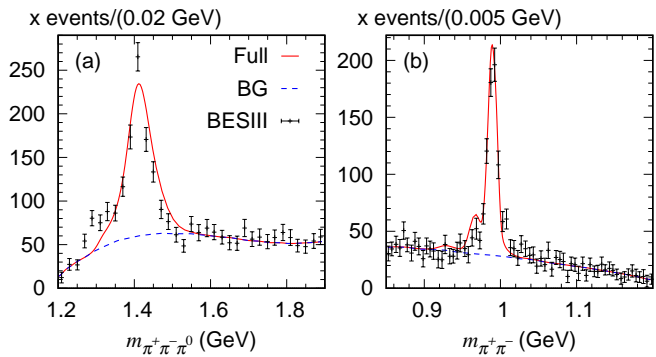}
\end{center}
 \caption{
(a) The $m_{\pi^+\pi^-\pi^0}$ and (b) $m_{\pi^+\pi^-}$ distributions for 
$J/\psi\to\gamma(\pi^+\pi^-\pi^0)$.
Our full results have been smeared with the bin width, scaled to fit the
 BESIII data~\cite{bes3-3pi}, and augmented by 
the background polynomials (BG) of \cite{bes3-3pi}.
 }
\label{fig:3pi}
\end{figure}
For the isospin-violating $J/\psi\to\gamma\eta(1405/1475)\to\gamma(\pi\pi\pi)$, our
coupled-channel model predicts the $m_{\pi^+\pi^-\pi^0}$ and $m_{\pi^+\pi^-}$ distributions as
shown in Figs.~\ref{fig:3pi}(a) and \ref{fig:3pi}(b), respectively,
in good agreement with the BESIII data~\cite{bes3-3pi}.
The authors of Refs.~\cite{wu2012,wu2013,du2019} proposed that these processes are dominantly caused by the $K^*K\bar{K}$ triangle loop mechanisms [Fig.~\ref{fig:diag}(c)].
While the $K^*K^+K^-$ and $K^*K^0\bar{K}^0$ triangle loops cancel each other
almost exactly for the isospin symmetry, due to $m_{K^\pm}\ne m_{K^0}$, a significant isospin violation occurs in the small window of $2 m_{K^\pm}\ltap m_{\pi^+\pi^-}\ltap 2 m_{K^0}$.
The mechanism satisfies the kinematical condition to cause a triangle singularity at $m_{\pi^+\pi^-\pi^0}\sim 1410$~MeV and $m_{\pi^+\pi^-}\sim 2 m_{K}$, and the peaks appear as a result, as shown in Fig.~\ref{fig:3pi}.
This mechanism is required by the three-body unitarity and, thus, automatically included in our calculation.

The $K\bar{K}\pi$ and $\pi\pi\pi$ branching ratios in Refs.~\cite{pdg,bes3-3pi}
give ratios:
$\Gamma  [J/\psi\to\gamma\eta(1405/1475)\to\gamma(\pi^+\pi^-\pi^0)]/$
$\Gamma  [J/\psi\to\gamma\eta(1405/1475)\to\gamma(K\bar{K}\pi)]$
= 0.004--0.007
and 
$\Gamma  [J/\psi\to\gamma\eta(1405/1475)\to\gamma(\pi^0\pi^0\pi^0)]/$
$\Gamma  [J/\psi\to\gamma\eta(1405/1475)\to\gamma(K\bar{K}\pi)]$
= 0.002--0.003.
Our coupled-channel model predicts 0.0045--0.0047 and 0.0015--0.0016, respectively, in good agreement with the experimental ones.
These reasonable predictions for the $\pi\pi\eta$ and $\pi\pi\pi$ final states support
the model's dynamical content.

\begin{table}[b]
\renewcommand{\arraystretch}{1.6}
\tabcolsep=3.9mm
\caption{\label{tab:pole}
Pole positions ($E_{\eta^*}$) labeled by $\alpha$.
The mass and width are $M={\rm Re} [E_{\eta^*}]$ and $\Gamma=-2{\rm Im} [E_{\eta^*}]$, 
respectively.
The Riemann sheet (RS) of $E_{\eta^*}$ is specified by $(s_{K^*\bar{K}},s_{a_2(1320)\pi})$ where $s_{x}=p(u)$ indicates the physical
 (unphysical) sheet of a channel $x$.
The BESIII result is Breit-Wigner parameters.
All errors are statistical.
}
\begin{tabular}{cccc}
            &  $M$ (MeV) &  $\Gamma$ (MeV) & RS \\\hline
 $\alpha=1$ & $1401.6      \pm       0.6$ & $65.8  \pm       1.0$ & $(up)$\\
 $\alpha=2$ & $1401.6      \pm       0.6$ & $66.3  \pm       0.9$ & $(pp)$\\
 $\alpha=3$ & $1495.0      \pm       1.5$ & $86.4  \pm       1.8$ & $(up)$\\\hline
\multirow{2}{*}{BESIII~\cite{bes3_mc}}
 & $1391.7\pm 0.7$ & $60.8\pm 1.2$ \\
                      & $1507.6\pm 1.6$ & $115.8\pm 2.4$ & 
\end{tabular}
\end{table}

\subsection{Pole structure of $\eta(1405/1475)$}

%
Extraction of poles from amplitudes that respect three-body unitarity
has long been discussed~\cite{gloeckle,flinders,julich-a1}.
However, until recently, this method had not been applied to data
involving three-body final states. 
A breakthrough was made by extracting an $a_1(1260)$ pole from
$m_{\pi^+\pi^-\pi^-}$ lineshape data for
$\tau^-\to\pi^+\pi^-\pi^-\nu_\tau$
with a $\rho\pi$ single-channel model~\cite{a1-jpac,a1-gwu}.
The three-body unitarity was rigorously (partially) considered in
Ref.~\cite{a1-gwu} (\cite{a1-jpac}).
Consequently,
an additional spurious pole was found in Ref.~\cite{a1-jpac},
indicating the importance of the full
three-body unitarity for studying pole structures. 
The analysis method of Ref.~\cite{a1-gwu}
should be further improved by 
considering coupled-channels and fitting Dalitz plots.
Below, we extract $\eta(1405/1475)$ poles with such improvements;
the three-body unitarity is treated as rigorously as done in
Ref.~\cite{a1-gwu}.

We search for $\eta(1405/1475)$ poles ($E_{\eta^*}$) that satisfy 
${\rm det}\,[\bar{G}^{-1}(E_{\eta^*})]=0$ 
with $\bar{G}^{-1}(E)$ defined in Eq.~(\ref{eq:mstar-g1}).
The analytic continuation of $\bar{G}^{-1}(E)$ involves 
appropriately deforming the integral paths in 
Eqs.~(\ref{eq:dressed-g}) and (\ref{eq:mstar-sigma})
to avoid crossing singularities and to select 
a relevant Riemann sheet (RS)~\cite{gloeckle,flinders,julich-a1,a1-gwu,SSL}.
Our analytic continuation method is quite similar to that described in Ref.~\cite{a1-gwu}.

\begin{figure}[t]
\begin{center}
\includegraphics[width=.49\textwidth]{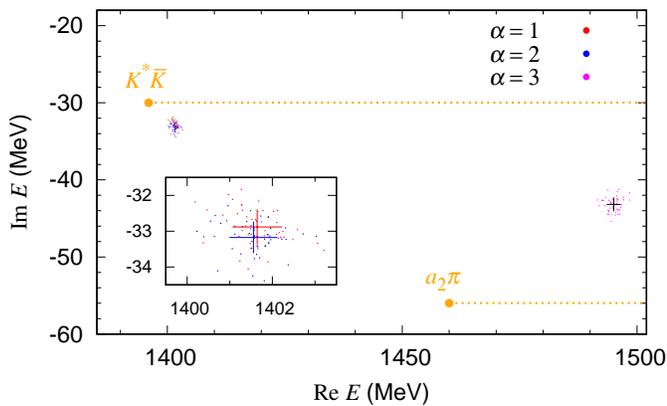}
\end{center}
 \caption{
Pole locations labeled by $\alpha$ from 50 bootstrap fits.
The crosses indicate averaged pole locations and their standard
 deviations.
The orange circles and dotted lines are the $K^*(892)\bar{K}$ and $a_2(1320)\pi$ branch
 points and cuts, respectively
The $\alpha=1,2$ region is enlarged in the inset.
}
\label{fig:pole}
\end{figure}

Three poles labeled by $\alpha=1,2$ [$\alpha=3$] corresponding to
$\eta(1405)$ [$\eta(1475)$] are found; see Table~\ref{tab:pole}.
Statistical errors are based on 50 bootstrap fits.
Although the mass and width values
from our analysis and the BESIII analysis (BW parameters) do not agree within the
errors, they are fairly similar.
The poles are close to the branch points associated with the $K^*(892)\bar{K}$ and $a_2(1320)\pi$ channels as shown in Fig.~\ref{fig:pole}.
Thus we specify the RS of these channels in Table~\ref{tab:pole}~\footnote{Section 50 of Ref.~\cite{pdg} defines the (un)physical sheet.
}.
The two-pole structure of $\eta(1405)$ does not mean two physical states but is simply due to the fact that a pole is split into two poles on different RSs of its decay channel.
The two $\eta(1405)$ pole values are very similar except for the RS 
due to their proximity to the $K^*(892)\bar{K}$ threshold.
As a consequence of the unitarity, the pole structure ($\eta^*$ propagation) and the Dalitz plot distributions ($\eta^*$ decay mechanism) are connected by the common dynamics in our model but not in the BW model.

The bare states in our model are conceptually similar to states given by quark models or LQCD without two-hadron operators. 
The new BESIII data~\cite{bes3_mc} require two bare states.
The lighter one of $\sim 1.6$~GeV seems compatible with the excited $s\bar{s}$~\cite{pdg,barnes1997}.
The heavier bare mass can be in the range of 2--2.4~GeV to give comparable fits.
This state could be either of a second radial excitation of $\eta^{(\prime)}$, a hybrid~\cite{dudek2013}, a glueball~\cite{bali1993,morningstar1999,chen2006,richards2010,chen2111}, or their mixture.
The two bare states are mixed and dressed by continuum coupled-channels to form the $\eta(1405/1475)$ poles.

\section{Summary and Outlook}
We conducted a coupled-channel analysis of radiative $J/\psi$ decays via $\eta(1405/1475)$,
and addressed the long-standing $\eta(1405/1475)$ puzzles itemized in the Introduction.
Our three-body unitary coupled-channel model is reasonably fitted to the
$K_SK_S\pi^0$ Dalitz plot pseudodata samples generated with $J^{PC}=0^{-+}$
amplitude of the BESIII MC~\cite{bes3_mc}, and also to branching
fractions of $\eta\pi^+\pi^-$ and $\gamma\pi^+\pi^-$ final states
relative to that of $K\bar{K}\pi$.
The model predicts the different $\eta(1405/1475)$ lineshapes for the $\eta\pi^+\pi^-$ and 
$\pi\pi\pi$ final states in reasonable agreement with experimental results.
The model also predicts the branching fractions well for the isospin-violating $\pi\pi\pi$ final states and their narrow $f_0(980)$-like $\pi^+\pi^-$ lineshape;
the triangle singularity effect automatically included in our unitary model plays a crucial role.
Our model revealed the $\eta(1405/1475)$ pole structure for the first time.
Two poles on different Riemann sheets of the $K^*\bar{K}$ channel correspond to $\eta(1405)$,
and one pole for $\eta(1475)$.

Last but not least,
what we presented is the first-ever pole determination based on 
a manifestly three-body unitary coupled-channel framework applied to
experimental Dalitz plot distributions.
In the future, the present analysis for $0^{-+}$ should be further
extended to include more $J^{PC}$ to analyze the radiative $J/\psi$
decay data directly,  consistently addressing pole structures of $\eta(1405/1475)$, $f_1(1420)$, etc. with the unitary coupled-channel framework.
This development is important since the present analysis results could
have been
biased by the $0^{-+}$ components in the radiative $J/\psi$ decays
determined with simpler Breit-Wigner models~\cite{bes3_mc,bes_jpsi-gamma-eta-pipi,mark3_jpsi-gamma-eta-pipi,bes2_rhog,mark3_rhog}.

\begin{acknowledgments}
The authors thank the BESIII Collaboration for providing us with the MC outputs for our study.
The authors also thank 
Jozef J. Dudek, T.-S. Harry Lee, Bei-Jiang Liu, Xiao-Hai Liu, Toru Sato,
 Guo-Fa Xu, Qiang Zhao, and Bing-Song Zou for useful discussions.
This work is in part supported by the National Natural Science Foundation of China (NSFC) under Grants No.U2032103, U2032111, 11625523, 12175239, 12221005, 
and also by the National Key Research and Development Program of China under Contracts 2020YFA0406400.
\end{acknowledgments}



\end{document}